%% file: Article_jos.tex
\documentclass[10pt,letterpaper]{article}
\pdfoutput=1 

\input{Structure_jos.tex} 

\makeatletter
\normalsize
\divide\@tempdima\baselineskip
\@tempcnta=\@tempdima
\makeatother

\DeclareFixedFont\trfont{OT1}{phv}{b}{sc}{11}

\title{%
       \vspace{-1.5cm}
       \centering\boldmath\LARGE\bfseries%
       \begin{adjustwidth}{-0.25in}{-0.25in}
       \centering
       Josephson AC effect induced by weak gravitational field
       \end{adjustwidth}
       \bigskip
       }

\author{\textsc{Giovanni Alberto Ummarino}
\vspace{0.1em}}
\affil{%
\makebox[\textwidth][c]{Politecnico di Torino, Dipartimento di Scienza Applicata e Tecnologia, corso Duca degli Abruzzi 24, 10129 Torino, Italy}%
}
\affil{National Research Nuclear University MEPhI, Kashirskoe hwy 31, 115409
       Moscow, Russia%
\vspace{-0.025em}
}%
\affil{\href{mailto:giovanni.ummarino@polito.it}{\texttt{giovanni.ummarino@polito.it}}
       }

\smallskip

\author{\textsc{Antonio Gallerati}
\vspace{0.1em}
}
\affil{%
\makebox[\textwidth][c]{Politecnico di Torino, Dipartimento di Scienza Applicata e Tecnologia, corso Duca degli Abruzzi 24, 10129 Torino, Italy}
       }
\affil{Istituto Nazionale di Fisica Nucleare, Sezione di Torino, via Pietro
       Giuria 1, 10125 Torino, Italy%
\vspace{-0.025em}
       }%
\affil{\href{mailto:antonio.gallerati@polito.it}{\texttt{antonio.gallerati@polito.it}}
      }

\date{}

\begin{document}

\maketitle
\smallskip

\begin{abstract}
\noindent
In this paper we examine the possibility of a Josephson AC effect between two superconductors induced by the Earth's gravitational field, making use of the gravito-Maxwell formalism. The theoretical framework exploits the symmetry between the weak field expansion of the gravitational field and the standard Maxwell formulation, combined with the Josephson junction physics. We also suggest a suitable experimental setup, analysing also the related possible difficulties in measurements.
\end{abstract}

\bigskip



\section{Introduction} \label{sec:Intro}
\sloppy
The simplest manifestation of the Josephson effect \cite{Barone1982physics} is represented by a circuit closed on a {superconductor-insulator-superconductor} (SIS) junction to which a constant potential difference $\Delta V$ is applied. The voltage, in turn, produces a sinusoidal superconductive current across the junction with pulsation $\omega={2\,e\,\Delta V}/{\hbar}$. In the following, we use this formulation to highlight a possible interaction between a superconductive condensate and the gravitational field.
Our idea stems from the remarkable phenomenon known as gravity-induced quantum interference, where the Earth gravitational potential is introduced in the Schrodinger equation \cite{Sakurai:2011zz}. The effect can be measured splitting a nearly monoenergetic beam of thermal neutrons and then considering the produced interference paths: a gravity-induced quantum mechanical phase shift is observed, due to the presence of the Earth's gravitational field \cite{Colella:1975dq}. The experiment shows quantum effects originating from the interaction of quantum particles with a classical, weak-field gravitational background. As in this case, measurable effects can manifest themselves in different physical situations; in particular, the SIS Josephson junction is proposed here as a promising framework for observing new experimental evidence.\par
For over half a century, the possible interplay between gravity and superconductivity has been investigated, starting with DeWitt's seminal work
\cite{DeWitt:1966yi}, continuing with the Podkletnov's pioneering, controversial experiment \cite{podkletnov1992possibility} and the theoretical work of Modanese, who first clarify the possible origin of the gravity/condensate interaction \cite{Modanese:1995tx}. In the same period there had been both theoretical \cite{schiff1966gravitation} and experimental \cite{witteborn1967experimental,witteborn1968experiments} works about generalized electric-type fields induced in metals by the presence of a gravitational field%
\footnote{%
for an updated summary on the subject, see \cite{Ummarino:2019cvw,Ummarino:2017bvz} and references therein.
}%
. The fundamental result of those researches was the introduction of a fundamental, generalized field of the form $\E=\Ee+\frac{m}{e}\:\Eg$, sum of an electrical component $\Ee$ and a gravitational one $\Eg$, where $m$ and $e$ are the electron mass and charge.\par
Another way to achieve the same results is the use the gravito-Maxwell formalism \cite{Ummarino:2017bvz,Ummarino:2019cvw}, that we briefly develop in the following section.


\section{Weak field approximation and generalized Maxwell equations}%
\label{sec:Maxeq}
Let us consider an almost flat spacetime configuration (weak gravitational field) where the metric $\gmetr$ and its inverse $\invgmetr$ can be expanded as:
\begin{equation}
\gmetr~\simeq~\emetr+\hmetr\;,\qqquad \invgmetr~\simeq~\invemetr-\invhmetr\;,
\label{eq:gmetr}
\end{equation}
$\hmetr$ being a small perturbation of the flat $\gmetr$ Minkowski metric%
\footnote{%
here we work in the mostly plus convention, {$\emetr=\mathrm{diag}(-1,+1,+1,+1)$},
and natural units $c=1$.}%
. The Christoffel symbols, to linear order in $\hmetr$, are written as
\begin{equation}\label{eq:Gam}
\Gam[\lambda][\mu][\nu]\=
    \frac12\,\invgmetr[\lambda][\rho]
    \left(\dm[\mu]\gmetr[\nu][\rho]+\dm[\nu]\gmetr[\rho][\mu]-\dm[\rho]\gmetr\right)
    \:\simeq\:\frac12\,\invemetr[\lambda][\rho]\,
     \left(\dm[\mu]\hmetr[\nu][\rho]+\dm[\nu]\hmetr[\rho][\mu]-\dm[\rho]\hmetr\right)\;.
\end{equation}
The Riemann tensor is defined as
\begingroup
\setlength{\belowdisplayskip}{8pt plus 2pt minus 2pt}%
\begin{equation}
\Ruddd\= 2\,\dm[{[}\lambda]\Gam[\sigma][\nu{]}][\mu]
        \+2\,\Gam[\sigma][\rho][{[}\lambda]\,\Gam[\rho][\nu{]}][\mu]\;,
\end{equation}
\endgroup
while the Ricci tensor is given by the contraction $\Ricci=\Ruddd[\sigma][\mu][\sigma][\nu]$ and, to linear order in $\hmetr$, it reads \cite{Wald:1984rg,Ummarino:2017bvz,Ummarino:2019cvw}
\begin{equation}
\Ricci~\simeq~\dmup[\rho]\dm[{(}\mu]\hmetr[\nu{)}][\rho]-\frac12\,\dd^2\hmetr-\frac12\,\dm\dm[\nu]h\;,
\qqquad h=\hud[\sigma][\sigma]\;.
\label{eq:Ricci}
\end{equation}
The Einstein equations \cite{Wald:1984rg,Misner:1974qy} are written as
\begin{equation}
\GEinst\:\equiv\:\Ricci-\dfrac12\,\gmetr\,R\=8\pi\GN\;T_{\mu\nu}\;,
\end{equation}
and the l.h.s.\ in first-order approximation reads \cite{Wald:1984rg,Ummarino:2017bvz,Ummarino:2019cvw}
\begin{equation}
\GEinst~\simeq~
    \dmup[\rho]\dm[{(}\mu]\bhmetr[\nu{)}][\rho]-\frac12\,\dd^2\bhmetr
    -\frac12\,\emetr\,\dmup[\rho]\dmup[\sigma]\bhmetr[\rho][\sigma]
    \=\dmup[\rho]\left(
          \dm[{[}\nu]\bhmetr[\rho{]}][\mu]+\dmup[\sigma]\emetr[\mu][{[}\rho]\,\bhmetr[\nu{]}][\sigma]
              \right)\;,
\end{equation}
having introduced the symmetric tensor
\begingroup%
\setlength{\abovedisplayshortskip}{2pt plus 3pt}%
\setlength{\abovedisplayskip}{2pt plus 3pt minus 4pt}
\setlength{\belowdisplayshortskip}{5pt plus 3pt}%
\setlength{\belowdisplayskip}{6pt plus 3pt minus 4pt}%
\begin{equation}
\bhmetr\=\hmetr-\frac12\,\emetr\,h\;.
\end{equation}
\endgroup%
If we also define the tensor \cite{Ummarino:2017bvz,Ummarino:2019cvw}
\begin{equation} \label{eq:Gscr}
\Gscr~\equiv~
\dm[{[}\nu]\bhmetr[\rho{]}][\mu]+\dmup[\sigma]\emetr[\mu][{[}\rho]\,\bhmetr[\nu{]}][\sigma]\;,
\end{equation}
the Einstein equations can be rewritten in the compact form:
\begin{equation}\label{eq:Einst}
\;\GEinst\=\dmup[\rho]\Gscr\=8\pi\GN\;T_{\mu\nu}\;.
\end{equation}
We can impose a gauge fixing using the harmonic coordinate condition (De Donder gauge) \cite{Wald:1984rg}
\begin{equation}
\Box x^\mu=0
\;\qLrq\;
\dm\left(\sqrt{-g}\,\invgmetr\right)=0
\;\qLrq\;
\invgmetr\,\Gam\,=\,0\;,
\label{eq:gaugefix}
\end{equation}
that, using eqs.\ \eqref{eq:gmetr} and \eqref{eq:Gam}, in turn implies the \emph{Lorentz gauge condition}
\begin{equation}
\dmup\bhmetr~\simeq~0\;.
\end{equation}
The latter simplifies the expression for $\Gscr$ in
\begingroup%
\setlength{\abovedisplayshortskip}{2pt plus 3pt}%
\begin{equation}
\Gscr~\simeq~\dm[{[}\nu]\bhmetr[\rho{]}][\mu]\;.
\label{eq:Gscr0}
\end{equation}
\endgroup

\paragraph{Gravito-Maxwell equations.}
Now, let us introduce the fields \cite{Ummarino:2017bvz,Ummarino:2019cvw}
\begin{align}
\Eg\equiv
    E_i=-\frac12\,\Gscr[0][0][i]=-\,\frac12\,\dm[{[}0]\bhmetr[i{]}][0]\;,\qquad
\Ag\equiv A_i=\frac14\,\bhmetr[0][i]\;,\qquad
\Bg\equiv B_i=\frac14\,{\varepsilon_i}^{jk}\,\Gscr[0][j][k]\;,
\end{align}
If we consider the divergence and curl of the above quantities, we obtain, restoring physical units, the set of equations \cite{Ummarino:2017bvz,Ummarino:2019cvw}:
\begingroup%
\setlength{\abovedisplayshortskip}{2pt plus 3pt}%
\setlength{\abovedisplayskip}{6pt plus 3pt minus 4pt}%
\begin{equation} \label{eq:gravMaxwell}
\begin{split}
&\nabla\cdot\Eg\=4\pi\GN\,\rhog\;;\\[2\jot]
&\nabla\cdot\Bg\=0 \;;\\[2\jot]
&\nabla\times\Eg~=-\dfrac{\dd\Bg}{\dd t} \;;\\[2\jot]
&\nabla\times\Bg\=4\pi\GN\,\frac{1}{c^2}\,\jg
                  \+\frac{1}{c^2}\,\frac{\dd\Eg}{\dd t}\;,
\end{split}
\end{equation}
\endgroup
where we have used \eqref{eq:Gscr0}, \eqref{eq:Einst} and introducing the mass density \:$\rhog\equiv-T_{00}$\: and the mass current density vector \:$\jg \equiv j_i \equiv T_{0i}$\,.
The above equations are formally equivalent to Maxwell equations, with $\Eg$ and $\Bg$ gravitoelectric and gravitomagnetic field respectively. For example, on the Earth surface, $\Eg$ is simply the Newtonian gravitational acceleration, while $\Bg$ is related to angular momentum interactions \cite{Braginsky:1976rb,Peng1983calculation,Peng1990approach}.\par\smallskip

\paragraph{Generalized Maxwell equations.}
Now let us introduce the generalized electric/magnetic field, scalar and vector potentials, containing both electromagnetic and gravitational terms:
\begin{equation}
\E=\Ee+\frac{m}{e}\,\Eg\,;\qquad
\B=\Be+\frac{m}{e}\,\Bg\,;\qquad
V=V_\text{e}+\frac{m}{e}\,V_\text{g}\,;\qquad
\A=\Ae+\frac{m}{e}\,\Ag\,,\quad
\label{eq:genfields}
\end{equation}
where $m$ and $e$ are the mass and electronic charge, respectively, the subscripts identifying the electromagnetic and gravitational contributions.\par
The generalized Maxwell equations for the fields \eqref{eq:genfields} then become \cite{Ummarino:2017bvz,Ummarino:2019cvw,Behera:2017voq}:
\begingroup%
\setlength{\abovedisplayshortskip}{2pt plus 3pt}%
\setlength{\abovedisplayskip}{5pt plus 3pt minus 4pt}%
\begin{equation} \label{eq:genMaxwell}
\begin{split}
&\nabla\cdot\E\=\left(\frac1\epsg+\frac{1}{\epsz}\right)\,\rho \;;\\[2\jot]
&\nabla\cdot\B\=0 \;;\\[2\jot]
&\nabla\times\E~=-\dfrac{\dd\B}{\dd t} \;;\\[2\jot]
&\nabla\times\B\=\left(\mug+\muz\right)\,\jj
                  \+\frac{1}{c^2}\,\dfrac{\dd\E}{\dd t} \;,
\end{split}
\end{equation}
\endgroup%
with $\epsz$ and $\muz$ electric permittivity and magnetic permeability in the vacuum, $\rho$ and $\jj$ electric charge density and electric current density, respectively, and writing the mass density and the mass current density vector as
%
\begin{equation}
\rhog\=\frac{m}{e}\,\rho\;,\qqquad
\jg\=\frac{m}{e}\;\jj\;,
\end{equation}
%
while defining the vacuum gravitational permittivity and the vacuum gravitational permeability as
\begingroup
\setlength{\belowdisplayskip}{20pt plus 2pt minus 2pt}%
\begin{equation}
\epsg=\frac{1}{4\pi\GN}\,\frac{e^2}{m^2}\;,\qqquad
\mug=\frac{4\pi\GN}{c^2}\,\frac{m^2}{e^2}\;.
\end{equation}
\endgroup
We have shown how to define a new set of generalized Maxwell equations for
generalized electric $\E$ and magnetic $\B$ fields, in the limit of weak gravitational field; similar strategies, for example, can be used to conjecture a gravitational-analogue Aharonov-Bohm electric effect \cite{ludwin2010gravitational}.
In the following sections, we are going to use these results to study the behaviour of a suitable oriented Josephson junction.


\section{Josephson effect induced by gravity}
If two superconductors are put in contact and the critical current in the contact region is much lower than that of the individual constituents, the configuration is called \emph{weak link}. The superconductors, if taken separately, have phase $\phi_{1,2}$ and amplitude $|\psi_{1,2}|$, with wave functions
\begin{equation}
\psi_{1,2}\=|\psi_{1,2}|\,\exp\left(i\,\phi_{1,2}\right)\=\sqrt{\rho_{1,2}}\;\exp\left(i\,\phi_{1,2}\right)\:,
\end{equation}
where $\rho_{1,2}$ are the probability amplitudes of Cooper pair densities.
Once the weak link is formed, coherence is established across the barrier, with a phase difference $\Delta\phi=\phi_{2}-\phi_{1}$ causing interference between the previously independent wavefunctions, so that the system can be described with a single wavefunction as a whole.\par
\sloppy
A typical realization of a weak link is a SIS tunnel junction, consisting of two
superconductors (that we take equal for simplicity) of thickness $L$ and {surface $A$}, separated by a very thin oxide layer of thickness ${\ell\ll L}$.
The evolution of the system can be then described using the time dependent Schrodinger equation
\begin{equation}
i\hbar\,\frac{\partial\psi}{\partial t}\=\mathcal{E}\,\psi\;.
\end{equation}
Once the contact is created, we can state that there exists an overlap between the left $\psi_{1}$ and the right $\psi_{2}$ wavefunctions. This means that an additional term must be added to take into account this wavefunctions' interaction, so that the rate of change of $\psi_{1}$ is proportional to the amount of the coupling to $\psi_{2}$ in left side. Similarly, a symmetric equation must also exist for the change rate of $\psi_{2}$, so that one globally has \cite{Feynman1965Feynman}
\begin{subequations}\label{eq:changerate}
\begin{align}
i\hbar\,\frac{\partial\psi_{1}}{\partial t} \=\mathcal{E}_{1}\,\psi_{1}+K\,\psi_{2}\;,
\label{eq:changerate1}
\\
i\hbar\,\frac{\partial\psi_{2}}{\partial t}\= \mathcal{E}_{2}\,\psi_{2}+K\,\psi_{1}\;.
\label{eq:changerate2}
\end{align}
\end{subequations}
If the superconductors are of the same type (SIS case) the probability amplitudes of Cooper pair densities are equal, $\rho_{1}=\rho_{2}=\rho$.
The quantities $\mathcal{E}_{1}$ and $\mathcal{E}_{2}$ are the ground state energies of the unperturbed system when $K=0$. The relevant quantity of course is just $\Delta\mathcal{E}=\mathcal{E}_{2}-\mathcal{E}_{1}$ and we choose the zero of energy to be halfway between $\mathcal{E}_{1}$ and $\mathcal{E}_{2}$.\par
Now, we can insert in \eqref{eq:changerate} the wavefunctions
$\psi_{1,2}=\sqrt{\rho}\,\exp(i\,\phi_{1,2})$, where each function is assumed to have a well-defined macroscopic phase, constant in space, and a well defined Cooper pair density. If we separate the real and imaginary part, we can write the two following equations:
\begin{subequations} \label{eq:changerateReIm}
\begin{align}
\frac{\partial\gamma}{\partial t}\=\frac{\Delta\mathcal{E}}{\hbar}\=0\;,
\label{eq:changerateRe}
\\[\jot]
\frac{\partial\rho}{\partial t}
        \=\frac{2\,K}{\hbar}\,\rho\,\sin(\gamma)\;,
\label{eq:changerateIm}
\end{align}
\end{subequations}
where $\gamma=\phi_{2}-\phi_{1}$ and that hold in the absence of applied voltage of any kind (electric or gravitational-like)%
\footnote{%
equation \eqref{eq:changerateIm} is written in the standard Josephson formalism \cite{fossheim2005superconductivity} with a little abuse of notation: the $\rho$-density involved in the derivative on the l.h.s.\ refers to the density of superconducting current across the interface, while the density on the r.h.s.\ refers to the global density of the Cooper pairs in the system, that is conserved (constant) being the system in the superconductive state.
}.
The supercurrent across the contact has the form
\begin{equation}
J_\text{s}\:=\,-2\,e\,\frac{\partial\rho}{\partial t}
     \:=\,-\frac{4\,e\,K}{\hbar}\,\rho\,\sin(\gamma)\=J_{0}\,\sin(\gamma)\;,
\end{equation}
and it suggests to us that a supercurrent is driven across the thin layer separating the superconductors, depending on the superconducting phase difference across the barrier itself.\par
If now we apply a constant voltage $\Delta V$ across the junction, we find an oscillatory variation of phase difference. This is motivated analysing the time dependence of the phase difference in \eqref{eq:changerateRe}, that has to be modified in order to take into account the applied constant voltage:
\begin{equation}
\frac{\partial\gamma}{\partial t}\=\frac{2\,e\,\Delta V}{\hbar}\;,
\end{equation}
that, after integration, gives
\begin{equation}
\gamma(t)\=\gamma_{0}+\frac{2\,e\,\Delta V}{\hbar}t\;,
\end{equation}
where $\gamma_{0}$ is an integration constant.
Since the Josephson current depends on $\sin(\gamma(t))$, the supercurrent results in
\begin{equation}
J_\text{s}\=J_{0}\,\sin\left(\gamma_{0}+\frac{2\,e\,\Delta V}{\hbar}\,t\right)\;.
\end{equation}
The amplitude of the tunnelling current $I_{0}=J_{0}\,A$ is temperature dependent, of course just for $T<\Tc$, and is described by Ambegaokar–Baratoff formula \cite{Ambegaokar:1963zz1,Ambegaokar:1963zz2} that gives
\begin{equation}
I_{0}\=\frac{\pi\,\Delta_{\textsc{s}}(T)}{2\,e\,R_\textsc{n}}\,\tanh\left(\frac{\Delta_\textsc{s}(T)}{2\,k_\textsc{b}\,T}\right)\;,
\end{equation}
where $R_\textsc{n}$ is the junction resistance in the normal state and $\Delta_\textsc{s}(T)$ is the superconductive gap.
Thus, when a Josephson junction is biased into a finite voltage state, the AC
Josephson supercurrent flows across the junction, governed by the time evolution of the superconducting wave function phase.\par
The production of an AC signal from a DC voltage may be understood as the result of the energy conversion of electron pairs into photons \cite{Saxena2009proximity}.
Let us remark that, in DC Josephson effect, the Josephson tunnelling barrier
behaves as a weak superconducting link between the two superconductors. The
phases of the superconducting wavefunctions become locked and, as a result, the
two superconductors behave as a single coherent system. The junction then acts
as a global weak superconductor.\par
\begin{figure}[H]
\captionsetup{skip=5pt,belowskip=15pt,font=small,labelfont=small,format=hang}
\centering
\includegraphics[width=0.8\textwidth,keepaspectratio]{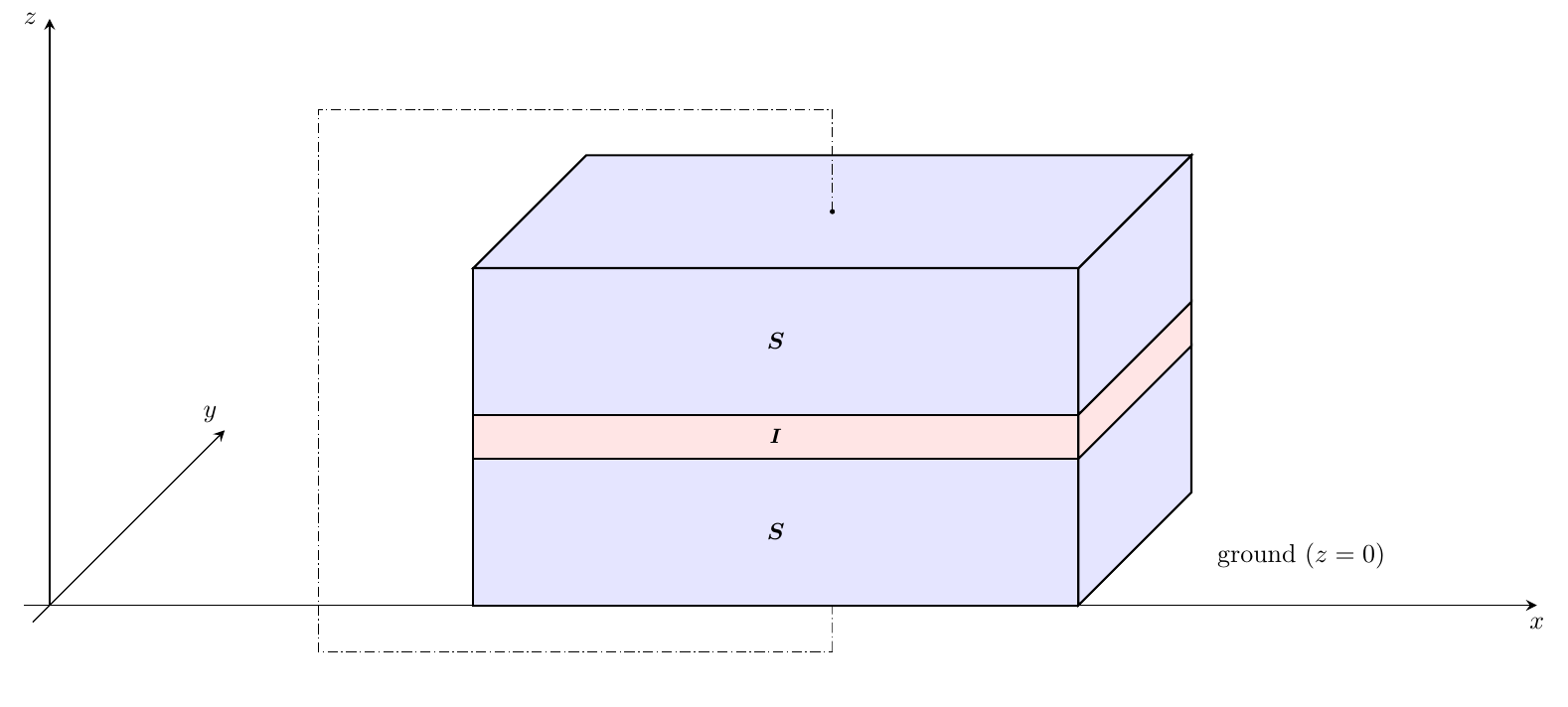}
\caption{SIS junction directed in the $z$ direction, the same of the Earth's gravitational field [color online].}
\label{fig:SIS}
\end{figure}

\smallskip
We have explained in Sect.\ \ref{sec:Maxeq} how a generalized electric field is expressed as $\E=\Ee+\frac{m}{e}\:\Eg$, while a generalized potential can be written in the form
$V=V_\text{e}\,+\,\frac{m}{e}\;V_\text{g}$. Now, if we restrict to the particular case
\begin{equation}
\Ee=0\;,\qquad \Eg=\textbf{g}\;,
\end{equation}
that is, a situation in which it is present only the Earth's gravitational field, we also have that
\begin{equation}
\Delta V\=\frac{m}{e}\,\Delta V_\text{g} \=\int_{0}^{\ell}\!dz\;\frac{m}{e}\,g\=\frac{m}{e}\,g\,\ell\;,
\end{equation}
since the gravitational field is directed along the $z$-axis (see Fig.\ \ref{fig:SIS}). The resulting Josephson current reads
\begin{equation}
I_\text{s}(t)\=I_0\,\sin\left(\frac{2\,e\,\Delta V}{\hbar}\,t+\varphi\right)\=I_0\,\sin(\omega\,t+\varphi)\;.
\end{equation}
Of course, if the junction is rotated so that the normal vector of the surface junction becomes perpendicular to the Earth's gravitational field, the effect disappears.\par
\begin{figure}[H]
\begin{adjustwidth}{-3.75em}{-3.75em}
\centering
\captionsetup[subfigure]{skip=5pt,belowskip=15pt,font=small,labelfont=small,format=hang}
\begin{subfigure}[t]{0.55\textwidth}
\includegraphics[width=\textwidth,keepaspectratio]{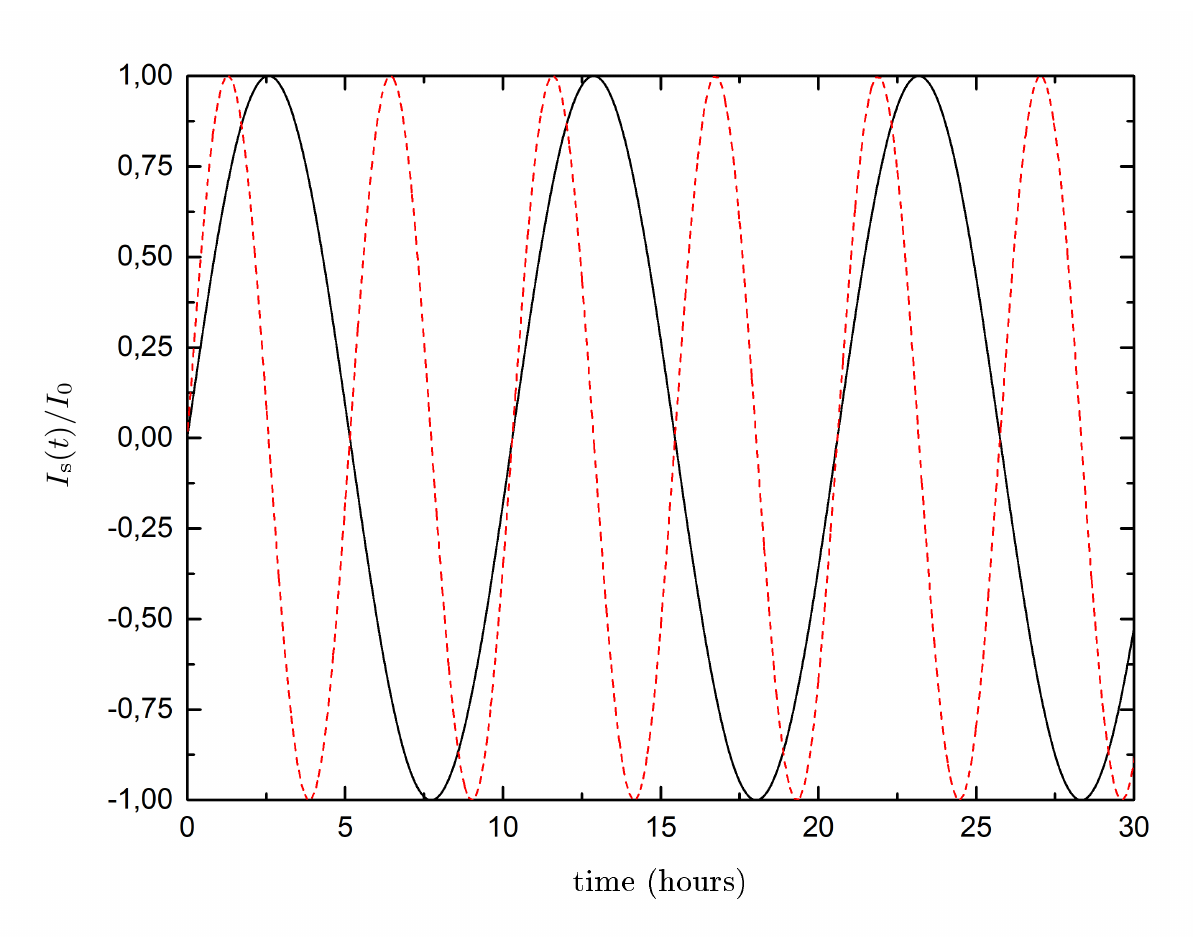}
\caption{Time dependence of the Josephson current for insulating layers of thickness $\ell=2\,\nm$ (red, dashed) and $\ell=1\,\nm$ (black, solid) [color online]. }
\label{subfig:Joscurr1}
\end{subfigure}
\hfill
\begin{subfigure}[t]{0.55\textwidth}
\includegraphics[width=\textwidth,keepaspectratio]{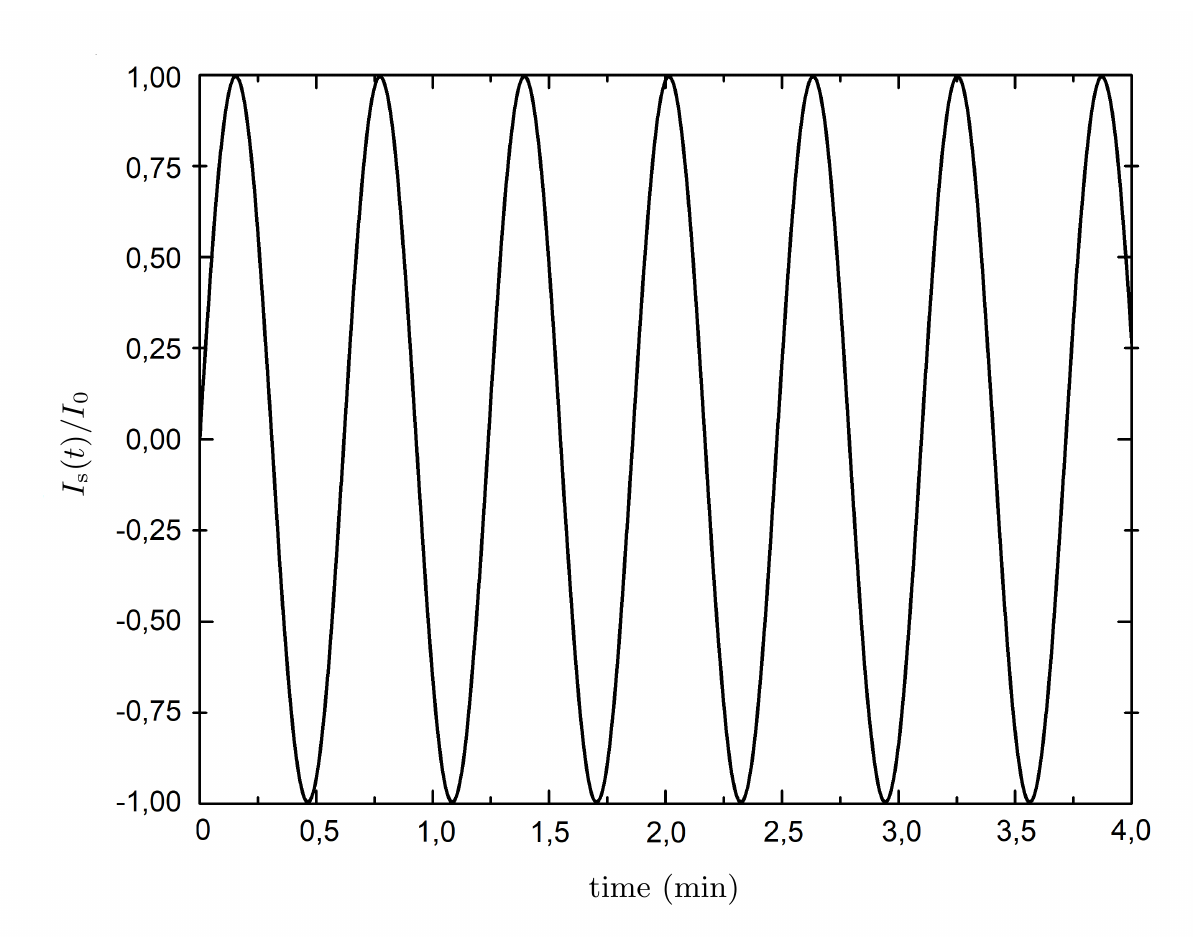}
\caption{Time dependence of the Josephson current for an insulating layer of thickness $\ell=1\,\mu\mathrm{m}$.}
\label{subfig:Joscurr2}
\end{subfigure}
\end{adjustwidth}
\end{figure}

\sloppy
The pulsation $\omega=\tfrac{2\,e\,\Delta V}{\hbar}$, for an insulating layer of thickness $\ell\in[1,\,2]\,\nm$, turns out to be in the range ${\omega\in[1.69\cdot10^{-4},\;3.38\cdot10^{-4}]\,\mathrm{s}^{-1}}$, and the corresponding oscillation period of the Josephson current results to be ${T=\tfrac{2 \pi}{\omega}\in[3.71\cdot10^4,\;1.85\cdot10^4]\,\mathrm{s}}$.
If one wants to observe the generated current, it is necessary to have a stable junction, since the duration of the experiment turns out to be approximately one day, see Fig.\ \subref{subfig:Joscurr1}.\par
If one wants to increase the voltage $V_\text{g}$, reducing this way the experiment duration, it is possible to use a junction made with Al. The latter becomes superconductive below $1.2\,\mathrm{K}$ and has a coherence length of $1.6\cdot10^{3}\,\nm$, so that we can take an insulating layer of thickness $\ell\simeq10^{3}\,\nm$. In this situation, we obtain for the voltage $V_\text{g}=5.6\cdot10^{-17}\,\mathrm{Volt}$, for the pulsation $\omega=1.69\cdot10^{-1}\,\mathrm{s}^{-1}$ and for the period $T=37.1\,\mathrm{s}$. This gives us an experiment duration of about 4 minutes, as shown in Fig.\ \subref{subfig:Joscurr2}.\par
From a practical point of view, the difficulty turns out to be the need for an experimental setup stable enough to allow accurate measurements. If we increase the junction thickness, the time duration for the experiment decreases, but the Josephson current becomes weaker. Probably, the best choice to observe clear experimental evidences is to realize the most stable setup possible, making long-time measurements.
\medskip


\section{Conclusions}
We have seen that the described theoretical model realizes the possibility to investigate the interplay between a superconductive condensate and gravitation. The difficulties lie in the experimental setup, that has to be stable in time, to allow for accurate long-duration observations, and very sensitive in the voltage. However, we think that experimental issues are not insurmountable and an effective verification of this idea should be possible.

\medskip

\medskip

%
%


\section*{\normalsize Acknowledgments}
\vspace{-5pt}
This work was supported by the MEPhI Academic Excellence Project (contract No.\ 02.a03.21.0005) for the contribution of prof.\ G.\ A.\ Ummarino.
We also thank Fondazione CRT \,\includegraphics[height=\fontcharht\font`\B]{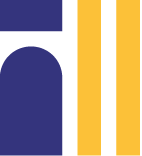}\:
that partially supported this work for dott.\ A.\ Gallerati.

%


\bigskip 
\hypersetup{linkcolor=blue}
\phantomsection 
\addtocontents{toc}{\protect\addvspace{3.5pt}}
\addcontentsline{toc}{section}{References} 
\bibliographystyle{mybibstyle}
\bibliography{bibliografia} 


\end{document}

%% file: Structure_jos.tex
\usepackage[utf8]{inputenc} 
\usepackage{lipsum} 

\usepackage[svgnames]{xcolor}  
\definecolor{Green}{rgb}{0.05, 0.45, 0.25}
\xdefinecolor{dogwoodrose}{rgb}{0.8, 0.1, 0.55}
\xdefinecolor{RRed}{rgb}{0.7, 0.1, 0.525}

\usepackage[top=2.5cm, bottom=3cm, left=3.5cm, right=3.5cm,
           heightrounded,marginparwidth=1.5cm, marginparsep=1cm]{geometry} 

\usepackage{changepage} 

\usepackage[shortlabels]{enumitem} 

\usepackage[square,numbers,merge,comma,sort&compress]{natbib} 
\makeatletter
\def\NAT@spacechar{\,}  
\makeatother

\usepackage{amsmath,amssymb,amsfonts,amsthm,amsbsy}
\usepackage{mathtools} 

\usepackage{fnpct}
\setfnpct{dont-mess-around} 

\usepackage{slashed,cancel}
\usepackage{comment}   
\usepackage{relsize}   
\usepackage{setspace}  
\usepackage{moresize}  
\usepackage{epsfig}
\usepackage{latexsym}
\usepackage{mathrsfs,calligra,aurical} 
\usepackage{calc}
\usepackage{float}
\usepackage{appendix}
\usepackage{xargs}
\usepackage{extarrows}
\usepackage{empheq}
\usepackage{soul} 

\usepackage{tikz}
\usetikzlibrary{scopes,decorations.pathmorphing,patterns,calc,arrows,
                shapes.geometric,shapes.arrows,decorations.markings,plotmarks}
\usepackage{pgfplots}

\usepackage[blocks]{authblk}  

\usepackage[fulladjust]{marginnote}%

\usepackage{graphicx} 

\usepackage[labelsep=colon]{caption}  
\usepackage[labelsep=colon,aboveskip=10pt,belowskip=10pt]{subcaption}
\usepackage{array,multirow,makecell,booktabs}  
\newcolumntype{M}[2]{>{\centering\arraybackslash$}#1{#2\linewidth}<{$}}
\newcolumntype{T}[2]{>{\centering\arraybackslash}#1{#2\linewidth}<{}}
\newcolumntype{R}[1]{>{\raggedleft\arraybackslash}m{#1\linewidth}<{}}
\newcolumntype{L}[1]{>{\raggedright\arraybackslash}m{#1\linewidth}<{}}

\makeatletter
\renewcommand\mcell@classz{\@classx
   \@tempcnta \count@
   \prepnext@tok
   \@addtopreamble{
      \ifcase\@chnum
         \hfil
         \mcell@agape{\d@llarbegin\insert@column\d@llarend}\hfil \or
         \hskip1sp
         \mcell@agape{\d@llarbegin\insert@column\d@llarend}\hfil \or
         \hfil\hskip1sp
         \mcell@agape{\d@llarbegin \insert@column\d@llarend}\or
         \mcell@agape{$\vcenter
         \@startpbox{\@nextchar}\insert@column\@endpbox$}\or
         \mcell@agape{\vtop
         \@startpbox{\@nextchar}\insert@column\@endpbox}\or
         \mcell@agape{\vbox
         \@startpbox{\@nextchar}\insert@column\@endpbox}%
      \fi
      \global\let\mcell@left\relax\global\let\mcell@right\relax
    }\prepnext@tok}
\makeatother

\usepackage{titlesec}

\titleformat{\section}{\normalfont\fontsize{11}{11}\bfseries}{\thesection}{0.5em}{}
\titleformat{\subsection}{\normalfont\normalsize\bfseries}{\thesubsection}{0.5em}{}
\titleformat{\subsubsection}{\normalfont\normalsize\bfseries}{\thesubsubsection}{0.5em}{}

\titlespacing*{\section}{0pt}%
                {4ex plus 1ex minus .5ex}{1.75ex plus .25ex minus .25ex}
\titlespacing*{\subsection}{0pt}%
                {3.5ex plus 1ex minus .5ex}{1.25ex plus .2ex minus .2ex}
\titlespacing*{\subsubsection}{0pt}%
                {2.5ex plus 0.75ex minus .2ex}{0.75ex plus .15ex minus .15ex}
\titlespacing*{\paragraph}{0pt}%
                {1.85ex plus 0.5ex minus .15ex}{1em}

\usepackage{titletoc}

\addtocontents{toc}{\addvspace{-0.75em}}  

\titlecontents{section}
  [1.25em] {\addvspace{0.7em plus 0pt}\small}
  {\thecontentslabel\hspace{0.75em}}{}
  {\hspace{0.5em}\titlerule*[0.5em]{.}\contentspage}
  [\addvspace{0.0em plus 0pt}]

\titlecontents{subsection}
  [2.75em] {\addvspace{0.075em plus0pt}\fns}
  {\thecontentslabel\hspace{0.75em}}{\thecontentslabel\hspace{0.75em}}
  {\hspace{0.5em}\titlerule*[0.5em]{.}\small\contentspage}
  [\addvspace{0.075em plus 0pt}]

\usepackage{environ}
\makeatletter
\NewEnviron{subalign}[1]{
\begin{subequations}\label{#1}
\begin{align} \BODY \end{align}
\end{subequations}      }
\makeatother
\makeatletter
{\begingroup%
\setlength{\abovedisplayskip}{10pt plus 4pt minus 9pt}%
\setlength{\abovedisplayshortskip}{0pt plus 2pt minus 2pt}%
\setlength{\belowdisplayskip}{12pt plus 3pt minus 9pt}%
\setlength{\belowdisplayshortskip}{7pt plus 3pt minus 4pt}%
\begin{subequations}%
}%
{\end{subequations}\ignorespacesafterend%
\endgroup}%
\makeatother

\makeatletter
{\begingroup%
\setlength{\abovedisplayskip}{{#1}pt plus 2pt minus 9pt}%
\setlength{\abovedisplayshortskip}{0pt plus 0pt minus 2pt}%
\setlength{\belowdisplayskip}{{#2}pt plus 3pt minus 9pt}%
\setlength{\belowdisplayshortskip}{7pt plus 3pt minus 4pt}%
\begin{subequations}%
}%
{\end{subequations}\ignorespacesafterend%
\endgroup}%
\makeatother

\makeatletter
{\begingroup%
\setlength{\abovedisplayskip}{{#1}pt plus 3pt minus 9pt}%
\setlength{\abovedisplayshortskip}{0pt plus 3pt}%
\setlength{\belowdisplayskip}{{#2}pt plus 3pt minus 9pt}%
\setlength{\belowdisplayshortskip}{7pt plus 3pt minus 4pt}%
\begin{equation}%
}%
{\end{equation}\ignorespacesafterend%
\endgroup}%
\makeatother

\makeatletter
{%
\begin{equation}%
\begin{split}%
}%
{\end{split}%
\end{equation}\ignorespacesafterend%
}%
\makeatother

\usepackage{bm}  
\usepackage{dsfont}  

\DeclareMathAlphabet{\mathpzc}{OT1}{pzc}{m}{it}
\DeclareMathAlphabet{\mathcal}{OMS}{cmsy}{m}{n}
\DeclareSymbolFontAlphabet{\Scr}{rsfs}
\DeclareMathAlphabet{\mathbold}{U}{BOONDOX-ds}{m}{n}
\SetMathAlphabet{\mathbold}{bold}{U}{BOONDOX-ds}{b}{n}
\DeclareMathAlphabet{\mathcalboondox}{U}{BOONDOX-calo}{m}{n}
\SetMathAlphabet{\mathcalboondox}{bold}{U}{BOONDOX-calo}{b}{n}
\DeclareMathAlphabet{\mathbcalboondox}{U}{BOONDOX-calo}{b}{n}

\newcommand\linkcol{RRed}


\usepackage[breaklinks=true,backref=page]{hyperref}
\hypersetup{
    bookmarks=true,         
    bookmarksnumbered=true,
    pdftoolbar=true,        
    pdfmenubar=true,        
    pdffitwindow=false,     
    pdfauthor={},     
    pdfsubject={},   
    pdfcreator={},   
    pdfproducer={},  
    pdfpagemode={UseNone},
    pdfstartview={FitH},
    colorlinks=true,
    plainpages,
    linktoc=page,
    citecolor=blue,
    filecolor=black,
    linkcolor=\linkcol,
    urlcolor=Green,
}
\renewcommand*{\backref}[1]{}
\renewcommand*{\backrefalt}[4]{%
\ifcase #1 %
\relax
\or
~{\small [\textsc{p.~\fns{\!#2}}]}
\else
~{\small [\textsc{p.~\fns{\!#2}}]}%
\fi}

\usepackage{footnotebackref}
\usepackage{hypernat} 

\def\+{~+~}
\def\-{~-~}
\def\={\:=\:}
\newcommand\fns{\footnotesize}

\newcommand\qqquad{\quad\quad\quad}

\newcommand\qLrq{\quad\Leftrightarrow\quad}

\newcommand\eps{\varepsilon}
\newcommand\epsz{\varepsilon_\ms{0}}
\newcommand\epsg{\varepsilon_\textrm{g}}
\newcommand\mug{\mu_\textrm{g}}
\newcommand\muz{\mu_\ms{0}}
\newcommand\w{\omega}

\newcommand\hgamma{\bar{\gamma}}

\newcommand\Tc{T_\textrm{c}}
\newcommand\E{\mathbf{E}}
\newcommand\B{\mathbf{B}}
\newcommand\A{\mathbf{A}}

\newcommand\Eg{\mathbf{E}_\textrm{g}}
\newcommand\Bg{\mathbf{B}_\textrm{g}}
\newcommand\Ag{\mathbf{A}_\textrm{g}}
\newcommand\Ee{\mathbf{E}_\textrm{e}}
\newcommand\Be{\mathbf{B}_\textrm{e}}
\newcommand\Ae{\mathbf{A}_\textrm{e}}
\newcommand\jj{\mathbf{j}}
\newcommand\jg{\mathbf{j}_\textrm{g}}

\newcommand\rhog{\rho_\textrm{g}}

\newcommand\nm{\mathrm{nm}}

\newcommand\GN{\mathrm{G}}

\newcommand{\ms}{\mathsmaller}

\newcommand{\dd}{\partial}
\newcommandx{\tts}[1]{\text{\textsmaller{#1}}}
\newcommandx{\dt}[1][1=f,usedefault]{\frac{\partial{#1}}{\partial t}}
\newcommandx{\dm}[1][1=\mu,usedefault]{\partial_{#1}}
\newcommandx{\dmup}[1][1=\mu,usedefault]{\partial^{#1}}
\newcommandx{\subm}[2][1=p,2=A,usedefault]{{#1}_{\!\mathsmaller{#2}}}
\newcommandx{\subt}[2][1=p,2=A,usedefault]{{#1}_\text{\textsmaller{#2}}}
\newcommandx{\supm}[2][1=p,2=A,usedefault]{{#1}^{\!\mathsmaller{#2}}}
\newcommandx{\supt}[2][1=p,2=A,usedefault]{{#1}^\text{\textsmaller{#2}}}
\newcommandx{\subpt}[3][1=p,2=A,3=B,usedefault]{{#1}^\text{\textsmaller{#3}}_\text{\textsmaller{#2}}}
\newcommandx{\subpm}[3][1=p,2=A,3=B,usedefault]{{#1}^{\mathsmaller{#3}}_{\mathsmaller{#2}}}
\newcommandx{\sh}[1][1=\alpha,usedefault]{\sinh\left(#1\right)}
\newcommandx{\ch}[1][1=\alpha,usedefault]{\cosh\left(#1\right)}
\newcommandx{\sech}[1][1=\alpha,usedefault]{\mathrm{sech}\left(#1\right)}
\newcommandx{\cosech}[1][1=\alpha,usedefault]{\mathrm{cosech}\left(#1\right)} \newcommandx{\LCTd}[4][1=\mu,2=\nu,3=\rho,4=\sigma,usedefault]{\eps_{#1#2#3#4}}
\newcommandx{\LCTu}[4][1=\mu,2=\nu,3=\rho,4=\sigma,usedefault]{\eps^{#1#2#3#4}}

\newcommandx{\gmetr}[2][1=\mu,2=\nu,usedefault]{g_{{#1}{#2}}}
\newcommandx{\invgmetr}[2][1=\mu,2=\nu,usedefault]{g^{{#1}{#2}}}
\newcommandx{\spc}[3][1=\mu,2=a,3=b,usedefault]{{\w_{#1}}^{\!\!{#2}{#3}}}
\newcommandx{\Conn}[3][1=\mu,2=\nu,3=\lambda,usedefault]{{\Gamma_{{#1}{#2}}}^{\!\!#3}}
\newcommandx{\viel}[2][1=\mu,2=a,usedefault]{{e_{#1}}^{\!#2}}
\newcommandx{\inviel}[2][1=a,2=\mu,usedefault]{{e_{#1}}^{#2}}
\newcommandx{\vieluu}[2][1=\mu,2=a,usedefault]{e^{#1#2}}
\newcommandx{\Rdduu}[4][1=\mu,2=\nu,3=a,4=b,usedefault]{{R_{{#1}{#2}}}^{{#3}{#4}}}
\newcommandx{\hgamui}[1][1=0,usedefault]{\hgamma^{\mathsmaller{#1}}}
\newcommandx{\hgamdi}[1][1=0,usedefault]{\hgamma_{{}_{#1}}}
\newcommandx{\gamui}[1][1=0,usedefault]{\gamma^{\mathsmaller{#1}}}
\newcommandx{\gamdi}[1][1=0,usedefault]{\gamma_{{}_{#1}}}

\newcommandx{\emetr}[2][1=\mu,2=\nu,usedefault]{\eta_{{#1}{#2}}}
\newcommandx{\invemetr}[2][1=\mu,2=\nu,usedefault]{\eta^{{#1}{#2}}}
\newcommandx{\hmetr}[2][1=\mu,2=\nu,usedefault]{h_{{#1}{#2}}}
\newcommandx{\invhmetr}[2][1=\mu,2=\nu,usedefault]{h^{{#1}{#2}}}
\newcommandx{\bhmetr}[2][1=\mu,2=\nu,usedefault]{\bar{h}_{{#1}{#2}}}
\newcommandx{\binvhmetr}[2][1=\mu,2=\nu,usedefault]{\bar{h}^{{#1}{#2}}}
\newcommandx{\hud}[2][1=\mu,2=\nu,usedefault]{{h^{#1}}_{\!\!#2}}
\newcommandx{\Ruddd}[4][1=\sigma,2=\mu,3=\lambda,4=\nu,usedefault]{{R^{#1}}_{\!{#2}{#3}{#4}}}
\newcommandx{\Gam}[3][1=\lambda,2=\mu,3=\nu,usedefault]{{\Gamma^{#1}}_{\!{#2}{#3}}}
\newcommandx{\Gamd}[3][1=\mu,2=\nu,3=\lambda,usedefault]{\Gamma_{{#1}{#2}{#3}}}
\newcommandx{\Ricci}[2][1=\mu,2=\nu,usedefault]{R_{{#1}{#2}}}
\newcommandx{\GEinst}[2][1=\mu,2=\nu,usedefault]{G^{{}^\tts{(E)}}_{{#1}{#2}}}
\newcommandx{\Gscr}[3][1=\mu,2=\nu,3=\rho,usedefault]{\mathscr{G}_{{#1}{#2}{#3}}}

%% file: Article_jos.bbl
\providecommand{\href}[2]{#2}\begingroup\begin{thebibliography}{10}

\bibitem{Barone1982physics}
Antonio Barone and Gianfranco Patern{\`{o}}, \textit{``{Physics and
  Applications of the Josephson Effect}''}; John Wiley {\&} Sons (1982).

\bibitem{Sakurai:2011zz}
Jun~John Sakurai and Jim Napolitano, \textit{``{Modern Quantum Mechanics}''};
  Cambridge University Press, Cambridge (2017).

\bibitem{Colella:1975dq}
R.~Colella, A.W. Overhauser and S.A. Werner, \textit{``{Observation of
  gravitationally induced quantum interference}''}, Phys. Rev. Lett.
  \textbf{34} (1975) 1472--1474.

\bibitem{DeWitt:1966yi}
Bryce~S. DeWitt, \textit{``{Superconductors and gravitational drag}''}, Phys.\
  Rev.\ Lett.\ \textbf{16} (1966) 1092--1093.

\bibitem{podkletnov1992possibility}
E.~Podkletnov and R.~Nieminen, \textit{``A possibility of gravitational force
  shielding by bulk \textrm{YBa${}_2$Cu${}_3$O${}_{7-\mathrm{X}}$}
  superconductor''}, Physica C: Superconductivity \textbf{203} (1992), n.~3-4,
  441--444.

\bibitem{Modanese:1995tx}
Giovanni Modanese, \textit{``{Theoretical analysis of a reported weak
  gravitational shielding effect}''}, Europhys.\ Lett. \textbf{35} (1996)
  413--418,
  [\href{http://arxiv.org/abs/hep-th/9505094}{\texttt{hep-th/9505094}}].

\bibitem{schiff1966gravitation}
L.I. Schiff and M.V. Barnhill, \textit{``Gravitation-induced electric field
  near a metal''}, Physical Review \textbf{151} (1966), n.~4, 1067.

\bibitem{witteborn1967experimental}
F.C. Witteborn and W.M. Fairbank, \textit{``Experimental comparison of the
  gravitational force on freely falling electrons and metallic electrons''},
  Physical Review Letters \textbf{19} (1967), n.~18, 1049.

\bibitem{witteborn1968experiments}
F.C. Witteborn and W.M. Fairbank, \textit{``Experiments to determine the force
  of gravity on single electrons and positrons''}, Nature \textbf{220} (1968),
  n.~5166, 436--440.

\bibitem{Ummarino:2019cvw}
Giovanni~Alberto Ummarino and Antonio Gallerati, \textit{``{Exploiting weak
  field gravity-Maxwell symmetry in superconductive fluctuations regime}''},
  Symmetry \textbf{11} (2019), n.~11, 1341,
  [\href{http://arxiv.org/abs/1910.13897}{\texttt{arXiv:1910.13897}}].

\bibitem{Ummarino:2017bvz}
Giovanni~Alberto Ummarino and Antonio Gallerati, \textit{``{Superconductor in a
  weak static gravitational field}''}, Eur. Phys. J. \textbf{C77} (2017), n.~8,
  549, [\href{http://arxiv.org/abs/1710.01267}{\texttt{arXiv:1710.01267}}].

\bibitem{Wald:1984rg}
Robert~M. Wald, \textit{``{General Relativity}''}; Chicago Univ. Pr., Chicago,
  USA (1984).

\bibitem{Misner:1974qy}
Charles~W. Misner, K.S. Thorne and J.A. Wheeler, \textit{``{Gravitation}''}; W.
  H. Freeman, San Francisco (1973).

\bibitem{Braginsky:1976rb}
Vladimir~B. Braginsky, Carlton~M. Caves and Kip~S. Thorne,
  \textit{``{Laboratory Experiments to Test Relativistic Gravity}''}, Phys.
  Rev. D \textbf{15} (1977) 2047.

\bibitem{Peng1983calculation}
Huei Peng, \textit{``On calculation of magnetic-type gravitation and
  experiments''}, General Relativity and Gravitation \textbf{15} (1983), n.~8,
  725--735.

\bibitem{Peng1990approach}
Huei Peng, \textit{``A new approach to studying local gravitomagnetic effects
  on a superconductor''}, General Relativity and Gravitation \textbf{22}
  (1990), n.~6, 609--617.

\bibitem{Behera:2017voq}
Harihar Behera, \textit{``{Comments on gravitoelectromagnetism of Ummarino and
  Gallerati in “Superconductor in a weak static gravitational field” vs
  other versions}''}, Eur. Phys. J. \textbf{C77} (2017), n.~12, 822,
  [\href{http://arxiv.org/abs/1709.04352}{\texttt{arXiv:1709.04352}}].

\bibitem{ludwin2010gravitational}
Doron~M. Ludwin, \textit{``{The gravitational analog of the Aharonov-Bohm
  electric effect}''}, arXiv preprint 1012.5603 (2010).

\bibitem{Feynman1965Feynman}
R.P. Feynman, R.B. Leighton and M.~Sands, \textit{``{The Josephson
  junction}''}, in {\em The Feynman Lectures on Physics}, vol.~III, ch.~21,
  sect.\ 21-9, Addison-Wesley Publ. Comp., New York (1965).

\bibitem{fossheim2005superconductivity}
Kristian Fossheim and Asle Sudb{\o}, \textit{``Superconductivity: physics and
  applications''}; John Wiley {\&} Sons Ltd (2004).

\bibitem{Ambegaokar:1963zz1}
Vinay Ambegaokar and Alexis Baratoff, \textit{``{Tunneling Between
  Superconductors}''}, Phys.\ Rev.\ Lett. \textbf{10} (1963) 486.

\bibitem{Ambegaokar:1963zz2}
Vinay Ambegaokar and Alexis Baratoff, \textit{``{Tunneling Between
  Superconductors (Errata)}''}, Phys.\ Rev.\ Lett. \textbf{11} (1963) 104.

\bibitem{Saxena2009proximity}
Ajay~Kumar Saxena, \textit{``{The Proximity and Josephson Effects}''}, in {\em
  High-Temperature Superconductors}, pp.~147--198, Springer Berlin Heidelberg
  (2009).

\end{thebibliography}\endgroup
